\documentclass[sigconf,natbib=true]{acmart}

\AtBeginDocument{%
  \providecommand\BibTeX{{%
    \normalfont B\kern-0.5em{\scshape i\kern-0.25em b}\kern-0.8em\TeX}}}

\copyrightyear{2023}
\acmYear{2023}

\usepackage{multicol}
\usepackage{multirow}
\usepackage{graphics}
\usepackage{xcolor}
\usepackage{algorithm}
\usepackage{algorithmicx}
\usepackage{overpic}
\usepackage{algpseudocode}
\usepackage{subcaption}
\usepackage{bbding}
\usepackage{enumitem}
\usepackage{balance}
\usepackage{hyperref}

\begin{document}

\title{Sequential Recommendation with Diffusion Models}


\author{Hanwen Du}
\email{hwdu@stu.suda.edu.cn}
\affiliation{%
\institution{Soochow University}
\city{Suzhou}
\state{Jiangsu}
\country{China}
}

\author{Huanhuan Yuan}
\email{hhyuan@stu.suda.edu.cn}
\affiliation{%
\institution{Soochow University}
\city{Suzhou}
\state{Jiangsu}
\country{China}
}

\author{Zhen Huang}
\email{zhuang02@stu.suda.edu.cn}
\affiliation{%
\institution{Soochow University}
\city{Suzhou}
\state{Jiangsu}
\country{China}
}

\author{Pengpeng Zhao$^{^*}$}
\thanks{$^*$Corresponding author.}
\email{ppzhao@suda.edu.cn}
\affiliation{%
\institution{Soochow University}
\city{Suzhou}
\state{Jiangsu}
\country{China}
}

\author{Xiaofang Zhou}
\email{zxf@cse.ust.hk}
\affiliation{%
\institution{The Hong Kong University of Science
and Technology}
\city{Hong Kong SAR}
\country{China}
}
\renewcommand{\shortauthors}{Hanwen Du et al.}

\begin{abstract}
Generative models, such as Variational Auto-Encoder (VAE) and Generative Adversarial Network (GAN), have been successfully applied in sequential recommendation. These methods require sampling from probability distributions and adopt auxiliary loss functions to optimize the model, which can capture the uncertainty of user behaviors and alleviate exposure bias. However, existing generative models still suffer from the posterior collapse problem or the model collapse problem, thus limiting their applications in sequential recommendation. To tackle the challenges mentioned above, we leverage a new paradigm of the generative models, i.e., diffusion models, and present sequential recommendation with diffusion models (DiffRec), which can avoid the issues of VAE- and GAN-based models and show better performance. While diffusion models are originally proposed to process continuous image data, we design an additional transition in the forward process together with a transition in the reverse process to enable the processing of the discrete recommendation data. We also design a different noising strategy that only noises the target item instead of the whole sequence, which is more suitable for sequential recommendation. Based on the modified diffusion process, we derive the objective function of our framework using a simplification technique and design a denoise sequential recommender to fulfill the objective function. As the lengthened diffusion steps substantially increase the time complexity, we propose an efficient training strategy and an efficient inference strategy to reduce training and inference cost and improve recommendation diversity. Extensive experiment results on three public benchmark datasets verify the effectiveness of our approach and show that DiffRec outperforms the state-of-the-art sequential recommendation models. \footnote{Our code will be available upon acceptance.}
 
\end{abstract}

\ccsdesc[500]{Information systems~Recommender systems}

\keywords{Sequential Recommendation; Diffusion Models; Generative Models}


\maketitle

\section{Introduction}
Sequential recommendation aims to infer the user's preferences from historical interaction sequence and predicts the next item that the user will probably take interest in the future. Traditional methods \cite{MDP,FPMC,Fossil} adopt Markov Chains (MCs) to capture the transition patterns between the items in the sequence. With the recent advancements in deep learning, various types of neural networks, such as Convolutional Neural Networks (CNNs) \cite{tang2018caser, NextItNet}, Recurrent Neural Networks (RNNs) \cite{srnn2016,Hidasi_2018} and Transformers \cite{kang18attentive,Sun2019bert} have been successfully applied in sequential recommendation. These methods represent items as a fixed-point vector and optimize model parameter via the Next Item Prediction (NIP) loss function. 

Despite effectiveness, these methods suffer from two major limitations. First, representing items as a fixed-point vector is a deterministic approach, which cannot well capture the uncertainty of user's behaviors \cite{STOSA, VSAN}. Second, adopting NIP as the only loss function may lead to exposure bias, especially when the data are sparse and noisy \cite{DebiasedBERT4Rec}. To address these issues, generative models such as Variational Auto-Encoder (VAE) \cite{VAE} and Generative Adversarial Network (GAN) \cite{GAN} have been successfully appied in sequential recommendation to model uncertainty and alleviate exposure bias. Generative models assume that the real data follow an unknown probability distribution and try to approximate that distribution using a neural network. For example, VSAN \cite{VSAN} applies variational inference to self-attention networks to model the uncertainty of user preferences, ACVAE \cite{ACVAE} enhances sequential VAE with adversarial training and contrastive loss function to capture more personalized and
salient characteristics of users, MFGAN \cite{MFGAN} adopts a Transformer-based generator and multiple factor-specific discriminators to train sequential recommenders in an adversarial style. 
Such schemes usually require sampling from probability distributions and then reparameterization using mean and variance learned from the generative models, which can well mimic the uncertainty of user behaviors. They also adopt auxiliary loss functions such as contrastive learning loss, Kullback-Leibler Divergence loss and discriminator loss, thus alleviating the exposure bias brought about by only using the NIP loss \cite{MFGAN}.

Despite effectiveness, existing generative models still suffer from two major problems, which limit their applications in sequential recommendation. First, both VAE and GAN train a neural network to generate the hidden representations of the sequences, but the quality of the generated hidden representations cannot be ensured due to the information bottleneck theory \cite{VIB}. As a result, the generated hidden representations may contain little information about the user's preferences, which is referred to as the posterior collapse problem \cite{posteriorcollapse}. Second, both VAE and GAN are prone to suffer from the model collapse problem \cite{ImproveGAN,modelcollapse}, yet introducing additional regularization techniques such as batch normalization \cite{batchnormalization} and label smoothing \cite{labelsmoothing} is not guaranteed to solve this problem. 

More recently, diffusion models \cite{NonequilibriumThermodynamics,DenoiseDiffusion} have emerged as a new paradigm of generative models. Inspired by nonequilibrium thermodynamics, diffusion models gradually convert the original data into Gaussian noises using a predefined Markov chain, which is called the \emph{forward process}. A neural network is then trained to reverse the noise in multiple diffusion steps so as to reconstruct original data from the noised data, which is called the \emph{reverse process}. As the forward process follows a predefined Markov chain without any learnable parameters, it is easy to control the quality of the generated hidden representations and avoid the posterior collapse problem. Moreover, instead of the complicated encoder-decoder or generator-discriminator structure in VAE or GAN, diffusion models only train a single neural network as data denoiser, which is much easier to optimize and less vulnerable to the model collapse problem. Therefore, diffusion models can avoid the issues of exisiting generative models such as VAE and GAN. 

While adopting diffusion models to remedy the problems of VAE or GAN seems a straightforward solution, directly applying diffusion models to sequential recommenders actually faces serious challenges. First, diffusion models were originally proposed to process data from continuous domains, such as image synthesis \cite{DiffusionBeatGAN} and audio synthesis \cite{DiffWave}. Recommender systems, by contrast, mostly process data from discrete domains, where each user or item is represented as a unique ID. Therefore, it remains an open challenge how to adapt the techniques designed for continuous domains to recommender systems with discrete data.
Second, the training and inference cost of diffusion models can be extremely huge, due to the lengthened diffusion steps required during training and inference. Suppose we set the diffusion steps as $N$, then during training and inference we need to calculate the output of the denoising neural network $N$ times so as to complete the reverse process, the time complexity of which is $N$ times more than VAE or GAN. As previous works \cite{DiffusionBeatGAN,DenoiseDiffusion} report that the diffusion steps required to achieve satisfactory performance is about $1000$, increasing the cost by $1000$ times can be unbearable for recommender systems, where the number of items and users can scale up to millions or even billions.


To tackle the challenges mentioned above, we present sequential recommendation with diffusion models (DiffRec), a diffusion-model based sequential recommender capable of making high-quality recommendations with efficient training and inference cost. Specifically, to adapt diffusion models for sequential recommendation, we define an additional transition in the forward process that maps the discrete items into its hidden representation, together with a transition in the reverse process that converts the hidden representation back to the probability distribution over the items. We also adopt a different noising strategy that only noises the target item instead of the whole sequence, which is more suitable for sequential recommendation. Next, based on the modified diffusion process, we derive the objective function of our framework and make it tractable with the help of a simplification technique that rewrites the KL-divergence term as the mean square error. As the simplified objective function requires training a neural network to predict the original data from the noised data, we design a denoise sequential recommender capable of denoising the hidden representations at different noise levels. Finally, to reduce the training and inference cost of diffusion models, we propose an efficient training that only samples the important diffusion steps instead of calculating the model output at all diffusion steps, which is able to achieve time complexity similar to most of the deep-learning based sequential recommenders. We also propose an efficient inference that is able to save computational cost and improve the diversity of recommendation result. Extensive experiments on three benchmark datasets verify the effectiveness of our approach. Our contributions are summarized as follows:
\begin{itemize}
[leftmargin =  8pt,topsep=1pt]
    \item We propose a diffusion-model based sequential recommender capable of making high-quality recommendations with efficient training and inference cost. To the best of our knowledge, we are the first to introduce diffusion models for sequential recommendation task.
    \item We design an additional transition in the forward and the reverse process of the diffusion models, together with a different noising strategy, to make diffusion models more suitable for sequential recommendation. Moreover, based on the modified diffusion process, we derive the objective function and design a denoise sequential recommender to fulfill the objective function. Finally, we propose an efficient training strategy and an efficient inference strategy to reduce training and inference cost and improve recommendation diversity.
    \item We conduct extensive experiments on three public benchmark datasets and show that DiffRec outperforms the state-of-the-art sequential recommendation models.
\end{itemize}
\section{RELATED WORK}
\subsection{Sequential Recommendation}
Early works on sequential recommendation adopt Markov Chains (MCs) to model the dynamic transition patterns between the interacted items in the sequence, such as MDP \cite{MDP}, FPMC \cite{FPMC} and Fossil \cite{Fossil}. With the advancements in deep learning, RNNs are applied to sequential recommendation to model user behavior sequences, such as GRU4Rec \cite{srnn2016,Hidasi_2018}, hierarchical RNN \cite{HRNN} and attention-enhanced RNN \cite{NARM}. CNNs are also shown effective in capturing sequential patterns, such as Caser \cite{tang2018caser} and NextItNet \cite{NextItNet}. Some works also enhance sequential recommendation with Graph Neural Networks (GNNs) in order to capture the complex transition patterns of items, such as SR-GNN \cite{SR-GNN} and GC-SAN \cite{GCSAN}. Recently, Transformers \cite{vaswani2017transformer} become the mainstream architecture for sequential recommendation owing to the flexibility of the self-attention mechanism, serving as the backbone of various sequential recommenders such as SASRec \cite{kang18attentive} and BERT4Rec \cite{Sun2019bert}. A more recent work STOSA \cite{STOSA} also replaces the standard dot product in the Transformer architecture with Wasserstein distance so as to model the uncertainty of user behaviors.

Recent years also witness a trend in applying self-supervised learning techniques to enhance sequential recommenders. Inspired by the success of data augmentation and contrastive learning in computer vision \cite{chen2020simple,mocov1,mocov2,mocov3} and natural language processing \cite{gao2021simcse,SCLlanguage,consert}, $\rm{S}^3$-Rec \cite{CIKM2020-S3Rec} designs self-supervised objectives via mutual information maximization to fuse attribute information into sequential recommendation, CL4SRec \cite{Xu2020Contrastive} proposes three data augmentation methods and a contrastive learning objective to alleviate the data sparsity issue for sequential recommendation, DuoRec \cite{DuoRec} designs both supervised and unsupervised sampling methods for contrastive learning in order to mitigate the representation degeneration problem in sequential recommendation. As this work focuses on how to design effective and efficient diffusion models for sequential recommendation, we leave the designing of the contrastive learning module for future work.

\subsection{Generative Models}
Generative models have been widely adopted in computer vision and natural language processing due to their powerful ability to approximate unknown probability distributions. Two representative paradigms, VAE \cite{VAE} and GAN \cite{GAN}, have been applied to various tasks, such as image synthesis \cite{VQVAE, brock2018large} and sentence generation \cite{SentenceVAE, SeqGAN}. More recently, diffusion models \cite{NonequilibriumThermodynamics,DenoiseDiffusion} have emerged as a new paradigm the generative models with a wide range of applications, such as controlled text generation \cite{li2022diffusion,gong2023sequence}, image generation \cite{latentdiffusion,DiffusionBeatGAN,Hierarchicaltext2image} and molecular generation \cite{EquivariantDiffusion, GeoDiff}.

In recommender systems, generative models have also been widely adopted to model the uncertainty of user behaviors and accurately infer the user's latent preferences. For example, MVAE \cite{VAE4CF} applies VAE to collaborative filtering with multinomial conditional likelihood, SVAE \cite{SVAE} combines VAE with RNN to model latent dependencies for sequential recommendation, ACVAE \cite{ACVAE} improves SVAE \cite{SVAE} by introducing adversarial and contrastive learning framework, IRGAN \cite{IRGAN} constructs a generative retrieval model and a discriminative retrieval model for information retrieval scenarios such as web search, item recommendation and question answering, RecGAN \cite{RecGAN} combines GAN with RNN to model both short-term and long-term user preferences for recommendation, MFGAN \cite{MFGAN} designs a Transformer-based sequence generator and multiple factor-specific
discriminators for sequential recommendation. Different from these works, our work applies diffusion models to generate items for sequential recommendation, which demonstrates better performance than VAE- or GAN-based methods.

\section{Background: Diffusion Models} \label{diffusionintroduction}
In this section, we provide a brief introduction on diffusion models. Assume we have an original data $\mathbf{h}^{0}\in\mathbb{R}^{d}$. In the \emph{forward process}, diffusion models gradually add noises to $\mathbf{h}^{0}$ through $N$ diffusion steps and yield a Markov chain $\mathbf{h}^{0}, \mathbf{h}^{1}, \cdots, \mathbf{h}^{n}, \cdots, \mathbf{h}^{N}$, until $\mathbf{h}^{N}$ is approximately Gaussian noise. Each transition $\mathbf{h}^{n-1} \rightarrow \mathbf{h}^{n}$ is defined as follows:
\begin{equation}
    q(\mathbf{h}^{n}|\mathbf{h}^{n-1}) = \mathcal{N} (\mathbf{h}^{n} ; \sqrt{1-\beta_{n}} \mathbf{h}^{n-1}, \beta_{n} \mathbf{I})
\end{equation}
where $\{\beta_{n}\}^{N}_{n=1}$ are a series of predefined noise schedules controlling the amount of noises added at each diffusion step, with $\beta_{n}\in (0,1)$. According to Ho et al. \cite{DenoiseDiffusion}, $\boldsymbol{h}^{n}$ can be directly obtained by the following equation:
\begin{equation}\label{gethn}
\begin{aligned}
q(\mathbf{h}^{n}|\mathbf{h}^{0}) &= \mathcal{N}(\mathbf{h}^{n}; \sqrt{\bar{\alpha}_n} \mathbf{h}^{0}, (1-\bar{\alpha}_n) \mathbf{I}) \\
    &= \sqrt{\bar{\alpha}_n} \mathbf{h}^{0} +  \sqrt{1-\bar{\alpha}_n}\epsilon,\thinspace\thinspace\thinspace \epsilon \sim \mathcal{N}(0, \mathbf{I}) 
    \end{aligned}
\end{equation}
where $\alpha_{n}=1-\beta_{n}$ and $\bar{\alpha}_n = \prod_{i=1}^n \alpha_i$ and $\bar{\alpha}_{N}{\approx}1$. This property means that we can directly obtain $\mathbf{h}_{t}$ at an arbitrary diffusion step given $\mathbf{h}_{0}$ by sampling from a random Gaussian noise $\epsilon$. Since $\bar{\alpha}_t$ is a predefined noise schedule without any learnable parameters, we can easily control the quality of $\mathbf{h}^{n}$ by defining an appropriate $\beta_{n}$. An investigation on the influences of different noise schedules is presented in Section~\ref{influencesofnoiseschedule}.

Ho et al. \cite{DenoiseDiffusion} further shows that the posterior Gaussian distribution $q(\mathbf{h}^{n-1}|\mathbf{h}^{n},\mathbf{h}^{0})$ can be made tractable given by:
\begin{equation}\label{getmean}
    \begin{aligned}
    q(\mathbf{h}^{n-1}|\mathbf{h}^{n},\mathbf{h}^{0}) &= \mathcal{N}(\mathbf{h}^{n-1}; \mu_{n}(\mathbf{h}^{n}, \mathbf{h}^{0}), \tilde{\beta}_n \mathbf{I}) \\
    \mu_{n}(\mathbf{h}^{n},\mathbf{h}^{0}) &\coloneqq \frac{\sqrt{\bar{\alpha}_{n-1}}\beta_n}{1-\bar{\alpha}_n}\mathbf{h}^{0} + \frac{\sqrt{\alpha_{n}}(1-\bar{\alpha}_{n-1})}{1-\bar{\alpha}_n} \mathbf{h}^{n}  \\
    \tilde{\beta}_{n} &\coloneqq \frac{1-\bar{\alpha}_{n-1}}{1-\bar{\alpha}_{n}} \beta_{n} 
    \end{aligned}
\end{equation}

Now that we have the noised data $\mathbf{h}^{n}$, we define a \emph{reverse process} that gradually converts $\mathbf{h}^{N}$ into the original data $\mathbf{h}^{0}$. Each transition $\mathbf{h}^{n} \rightarrow \mathbf{h}^{n-1}$ is defined as follows: 
\begin{equation}\label{pthetahn1hn}
    p_{\theta}(\mathbf{h}^{n-1}|\mathbf{h}^{n})=\mathcal{N}(\mathbf{h}^{n-1};\mu_{\theta}(\mathbf{h}^{n}, n), \Sigma_{\theta}(\mathbf{h}^{n}, n))
\end{equation}
where the learning of the mean $\mu_{\theta}$ is based on a neural network $f_{\theta}$, the variance $\Sigma_{\theta}=\beta_n\mathbf{I}$ is set as untrainable for stability \cite{DenoiseDiffusion}.

We can then use the Variational Lower Bound (VLB) \cite{NonequilibriumThermodynamics} as the objective function, which calculates the KL-divergence between $p_{\theta}$ and $q$ plus entropy terms:
\begin{equation}\label{VLB}
    \begin{aligned}
     &\mathcal{L}_{\rm{VLB}} = \mathbb{E}_{q(\mathbf{h}^{1:N} | \mathbf{h}^{0})}\left[\mathcal{L}_{N}  + \cdots + \mathcal{L}_{n} +\cdots + \mathcal{L}_{0}\right]\\
    &= \mathbb{E}_{q(\mathbf{h}^{1:N} | \mathbf{h}^{0})}\left[\underbrace{\log \frac{q(\mathbf{h}^{N} | \mathbf{h}^{0})}{p_{\theta}(\mathbf{h}^{N})}}_{\mathcal{L}_{N}} + \sum_{n=2}^N \underbrace{\log \frac{q(\mathbf{h}^{n-1} | \mathbf{h}^{n},\mathbf{h}^{0})} {p_{\theta}(\mathbf{h}^{n-1} | \mathbf{h}^{n}) }}_{\mathcal{L}_{n}} - \underbrace{{\log p_{\theta}( \mathbf{h}^{0} | \mathbf{h}^{1})}}_{\mathcal{L}_{0}}\right]    
    \end{aligned}
\end{equation}
\section{Proposed Framework}
In this section, we introduce our proposed framework, sequential recommendation with diffusion models (DiffRec) in detail. We start by formally defining the problem of sequential recommendation in Section \ref{statement}. Next, we introduce the modifications required to adapt diffusion models for sequential recommendation in Section \ref{modification}. Moreover, we introduce the derivation of the objective function in Section \ref{objectivefunction} and the architecture of the denoise sequential recommender in Section \ref{DSR}. Finally, we introduce the efficient training and inference strategy in Section \ref{trainingandinference}.
\subsection{Problem Statement}\label{statement}
In sequential recommendation, we denote $ \mathcal{U}$ as a set of users and $  \mathcal{V}$ as a set of items. For each user $u \in \mathcal{U}$, we sort the user's interacted items in the chronological order and construct a sequence $s_u=[v_1,v_2,\cdots\, v_t, \cdots,v_{T-1}, v_{T}]$, where $v_t\in \mathcal{V}$ denotes the item that user $u$ has interacted with at the $t$-th timestamp, $T$ denotes the maximum sequence length. To predict the next item that user $u$ is probably interested in, during training, a sequential recommender 
learns to maximize the probability of recommending the target item $v_{T}$ given the previously interacted items $[v_1,v_2,\cdots,v_{T-1}]$, which is denoted as $p({v_T|v_{1},v_{2},\cdots,v_{T-1}})$.
During inference, the sequential recommender predicts the probability of recommending the next item $v_{T+1}$ based on the whole sequence $[v_1,v_2,\cdots,v_{T}]$, which can be denoted as $p({v_{T+1}|v_{1},v_{2},\cdots,v_{T-1},v_{T}})$.

To denote the hidden representation of each item at different diffusion steps, we append the superscript $t$ to the notation $\mathbf{h}^{n}$ described in Section \ref{diffusionintroduction} and get $\mathbf{h}^{n}_{t}(1{\leq}n{\leq}N, 1{\leq}t{\leq}T)$, meaning the hidden representation of item $v_t$ at diffusion step $n$. The hidden representation of each item is concatenated together to construct a sequence, and we use $\mathbf{H}^{n}=[\mathbf{h}^{n}_{1},\mathbf{h}^{n}_{2},\cdots,\mathbf{h}^{n}_{t},\cdots,\mathbf{h}^{n}_{T}]\in\mathbb{R}^{T\times{d}}$ to denote the hidden representation of the sequence at diffusion step $n$. We will use this notation in the following passage.
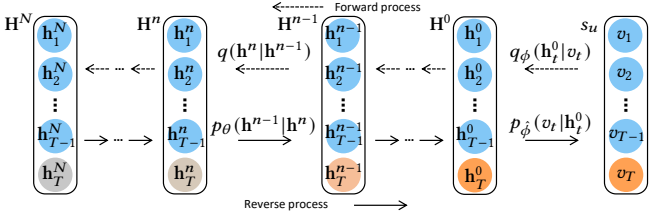
\begin{figure}
 \centering
 \footnotesize
 \begin{overpic}[width=0.46\textwidth]{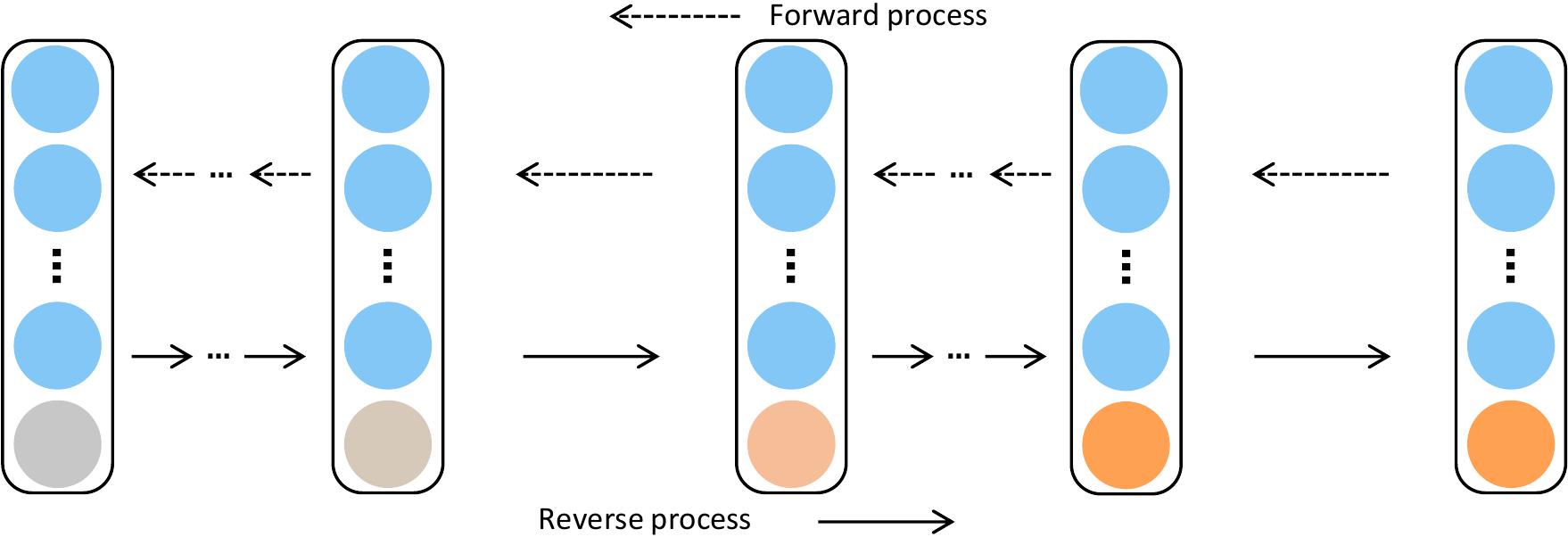}
\put(94.8,28){$v_1$}
\put(94.8,21.8){$v_2$}
\put(93.5,11.8){$v_{T-1}$}
\put(94.8,5.4){$v_{T}$}

\put(70.2,27.9){$\mathbf{h}^{0}_{1}$}
\put(70.2,21.5){$\mathbf{h}^{0}_{2}$}
\put(68.8,11.5){$\mathbf{h}^{0}_{T-1}$}
\put(70.2,4.8){$\mathbf{h}^{0}_{T}$}

\put(47.5,27.9){$\mathbf{h}^{n-1}_{1}$}
\put(47.5,21.5){$\mathbf{h}^{n-1}_{2}$}
\put(47.5,11.8){$\mathbf{h}^{n-1}_{T-1}$}
\put(47.5,5.2){$\mathbf{h}^{n-1}_{T}$}

\put(23,27.9){$\mathbf{h}^{n}_{1}$}
\put(23,21.5){$\mathbf{h}^{n}_{2}$}
\put(21.9,11.5){$\mathbf{h}^{n}_{T-1}$}
\put(23,5.2){$\mathbf{h}^{n}_{T}$}

\put(1.6,27.9){$\mathbf{h}^{N}_{1}$}
\put(1.6,21.5){$\mathbf{h}^{N}_{2}$}
\put(0.8,11.5){$\mathbf{h}^{N}_{T-1}$}
\put(1.6,5.2){$\mathbf{h}^{N}_{T}$}

\put(77.5,25){${q_{\phi}}(\mathbf{h}^{0}_{t}|v_t)$}
\put(77.5,13.5){${p_{\hat{\phi}}}(v_t|\mathbf{h}^{0}_{t})$}
\put(30,25){$q(\mathbf{h}^{n}|\mathbf{h}^{n-1})$}
\put(29,13){$p_{\theta}(\mathbf{h}^{n-1}|\mathbf{h}^{n})$}

\put(89,29.5){$s_u$}
\put(64.5,29.5){$\mathbf{H}^{0}$}
\put(39.8,29.5){$\mathbf{H}^{n-1}$}
\put(17,29.5){$\mathbf{H}^{n}$}
\put(-4.5,29.5){$\mathbf{H}^{N}$}
\end{overpic}
 \caption{An overview of the diffusion process. To adapt diffusion models for sequential recommendation, we define a transition $v_t\rightarrow\mathbf{h}^{0}_{t}$ in the forward process that converts the discrete item $v_t$ into its hidden representation $\mathbf{h}^{0}_{t}$, together with a transition $\mathbf{h}^{0}_{t}\rightarrow{v_t}$ in the reverse process converts the hidden representation back to the probability distribution over the items. Noising and denoising are only performed on the target item $v_T$.}\label{diffusionfigure}\end{figure}
\subsection{Adapt Diffusion Models for Sequential Recommendation}\label{modification}
Diffusion models are naturally defined to process continuous data, such as images and audios. To adapt diffusion models for sequential recommendation, in the forward process, we define an additional transition that maps the discrete item into its hidden representation. The hidden representation of the items has a continuous form, thus enabling the diffusion models to process the discrete recommendation data. Next, we modify the noising strategy in the forward process of the diffusion models that only noise the hidden representation of the target item instead of the whole sequence. Finally, in the reverse process, we accordingly define another transition that converts the hidden representations back to the probability over the items. An illustration of our modified diffusion process is presented in Figure \ref{diffusionfigure}.
\subsubsection{Forward Process}
To adapt diffusion models for sequential recommendation, we first define an additional transition $v_t\rightarrow\mathbf{h}^{0}_{t}$ that converts the discrete item $v_t$ into its hidden representation $\mathbf{h}^{0}_{t}$: 
\begin{equation}\label{qphiht0vt}
    {q_{\phi}}(\mathbf{h}^{0}_{t}|v_t)=\mathcal{N}(\mathbf{E}(v_t), \beta_0 \mathbf{I}),\thinspace\thinspace\thinspace \mathbf{E}(v_t)=\mathbf{v}_t
\end{equation}
where $\mathbf{E}\in\mathbb{R}^{|\mathcal{V}|\times{d}}$ is the embedding table with hidden dimensionality $d$, $\phi$ is the parameter of the embedding table, $\mathbf{v}_t\in\mathbb{R}^{d}$ is the embedding vector for item $v_t$. Empirically, we do not add noises at the embedding stage, so we set $\beta_0=0$.

Next, we modify the noising strategy in the forward process of the diffusion models to make it suitable for sequential recommendation. We decide not to noise the hidden representation of the whole sequence $\mathbf{H}^{n}$, but instead only noise the hidden representation of the target item $\mathbf{h}^{n}_{T}$. This is because the objective of a sequential recommender is to predict the target item, and noising the target item helps the sequential recommender obtain better predictions. In our experiments (Section~\ref{EffectivenessofNoisingStrategies}), we also find out that noising the hidden representation of the whole sequence may corrupt important information about the user's preference, which will hurt the performance of the sequential recommender. Therefore, we redefine Eq. \ref{gethn} as follows:
\begin{equation}\label{qhtnh0}
q(\mathbf{h}^{n}_{t}|\mathbf{h}^{0}_{t})=   \begin{cases}\mathbf{h}^{0}_{t}
  & \text{if}\ t<T \\
\sqrt{\bar{\alpha}_n} \mathbf{h}^{0}_{t} +  \sqrt{1-\bar{\alpha}_n}\epsilon,\thinspace\thinspace\thinspace \epsilon \sim \mathcal{N}(0, \mathbf{I}) & \text{if}\ t=T
\end{cases}
\end{equation}
\subsubsection{Reverse Process} Now that we have modified the forward process of our diffusion model, the reverse process should also be modified so as to match the forward process. Since we add a transition $v_t\rightarrow\mathbf{h}^{0}_{t}$ at the forward process, we accordingly add another transition $\mathbf{h}^{0}_{t}\rightarrow{v_t}$ that converts the hidden representation back to the probability distribution over the items:
\begin{equation}\label{pphihatvtht0}
        {p_{\hat{\phi}}}(v_t|\mathbf{h}^{0}_{t})=\rm{softmax}(\mathbf{W}\mathbf{h}^{0}_{t}+\mathbf{b})
\end{equation}
Here we use a linear layer with weight matrix $\mathbf{W}\in {\mathbb{R}}^{\mathcal{\lvert V \rvert}\times d}$ and bias matrix $\mathbf{b}\in \mathbb{R}^{\mathcal{\lvert V \rvert}}$, and $\hat{\phi}$ denote the parameters of this linear layer. This transition maps the continuous hidden representation $\mathbf{h}^{0}_{t}$ back to discrete items, and by calculating ${p_{\hat{\phi}}}(v_T|\mathbf{h}^{0}_{T})$ at the last timestamp $T$ we will know the probability of each item being recommended.

\subsection{Derive the Objective Function}\label{objectivefunction} We now introduce the derivation of the objective function for our framework. With the additional transition added to the diffusion model, we need to add another KL-divergence term between $p_{\hat{\phi}}$ and $q_{\phi}$ to Eq. \ref{VLB} so as to control the behavior of this transition:
\begin{equation}\label{VLB2}
    \begin{aligned}
    \mathcal{L}^{'}_{\rm{VLB}} &=\mathbb{E}_{q(\mathbf{h}^{0:N}_{T} | v_T)}\left[ \mathcal{L}_{\rm{VLB}}+\log\frac{{q_{\phi}}(\mathbf{h}^{0}_{T}|v_T)}{p_{\hat{\phi}}(v_T|\mathbf{h}^{0}_{T})}\right]\\
    &= \mathbb{E}_{q(\mathbf{h}^{0:N}_{T} | v_T)}\Bigg[ \log \frac{q(\mathbf{h}^{N}_{T} | \mathbf{h}^{0}_{T})}{p_{\theta}(\mathbf{h}^{N}_{T})}+ \sum_{n=2}^N \log \frac{q(\mathbf{h}^{n-1}_{T} | \mathbf{h}^{n}_{T},\mathbf{h}^{0}_{T})} {p_{\theta}(\mathbf{h}^{n-1}_{T} | \mathbf{h}^{n}_{T}) } \\&+ {\log\frac{{q_{\phi}}(\mathbf{h}^{0}_{T}|v_T)}{ p_{\theta}( \mathbf{h}^{0}_{T} | \mathbf{h}^{1}_{T})}}-\log{p_{\hat{\phi}}(v_T|\mathbf{h}^{0}_{T})}\Bigg]
    \end{aligned}
\end{equation}
Note that as we only add noise to the hidden representation of the target item $\mathbf{h}^{n}_{T}$, the VLB is therefore computed only on the hidden representation at the last timestamp $T$. 

Directly using Eq. \ref{VLB2} as the objective function is very difficult, because we have to train $N$ neural networks, each of which is in charge of reconstructing the data at one diffusion step. Luckily, Ho et al. \cite{DenoiseDiffusion} proposes a simplification technique that rewrites the KL-divergence term as the Mean Square Error (MSE) between the sampled mean $\mu_{n}$ and the predicted mean $\mu_{\theta}$:
\begin{equation}
\begin{aligned}\label{simple1}
&\mathbb{E}_{q(\mathbf{h}^{1:N}_{T} | \mathbf{h}^{0}_{T})} \left[ \log \frac{q(\mathbf{h}^{n-1}_{T} | \mathbf{h}^{n}_{T},\mathbf{h}^{0}_{T})} {p_\theta(\mathbf{h}^{n-1}_{T} | \mathbf{h}^{n}_{T}) }\right]\\ &= \mathbb{E}_{ q(\mathbf{h}^{1:N}_{T} | \mathbf{h}^{0}_{T})}  \left[ \frac{1}{2\Sigma_\theta^2}  || {\mu}_{n}(\mathbf{h}^{n}_{T}, \mathbf{h}^{0}_{T}) -\mu_\theta(\mathbf{h}^{n}_{T}, n)||^2\right] + C\\
&\propto \mathbb{E}_{ q(\mathbf{h}^{1:N}_{T} | \mathbf{h}^{0}_{T})} ||{\mu}_{n}(\mathbf{h}^{n}_{T}, \mathbf{h}^{0}_{T}) -\mu_\theta(\mathbf{h}^{n}_{T}, n) ||^2
\end{aligned}
\end{equation}
where ${\mu}_{n}(\mathbf{h}^{n}_{T}, \mathbf{h}^{0}_{T})$ is the sampled mean as defined in Eq. \ref{getmean} and the predicted mean $\mu_\theta(\mathbf{h}^{n}_{T})=\frac{\sqrt{\bar{\alpha}_{n-1}}\beta_n}{1-\bar{\alpha}_n}f_\theta(\mathbf{h}^{n}_{T},n) + \frac{\sqrt{\alpha_{n}}(1-\bar{\alpha}_{n-1})}{1-\bar{\alpha}_n} \mathbf{h}^{n}_{T}$ is derived in a similar way. We can further write:
\begin{equation}
    \begin{aligned}\label{simple2}
        & || \mu_\theta(\mathbf{h}^{n}_{T}, n) - {\mu}_{n}(\mathbf{h}^{n}_{T}, \mathbf{h}^{0}_{T}) ||^2\\
        =&||\frac{\sqrt{\bar{\alpha}_{n-1}}\beta_n}{1-\bar{\alpha}_n}\mathbf{h}^{0}_{T} + \frac{\sqrt{\alpha_{n}}(1-\bar{\alpha}_{n-1})}{1-\bar{\alpha}_n} \mathbf{h}^{n}_{T}\\&-\frac{\sqrt{\bar{\alpha}_{n-1}}\beta_n}{1-\bar{\alpha}_n}f_\theta(\mathbf{h}^{n}_{T},n) - \frac{\sqrt{\alpha_{n}}(1-\bar{\alpha}_{n-1})}{1-\bar{\alpha}_n} \mathbf{h}^{n}_{T}||^2\\
        =&\frac{\sqrt{\bar{\alpha}_{n-1}}\beta_n}{1-\bar{\alpha}_n}||\mathbf{h}^{0}_{T}-f_\theta(\mathbf{h}^{n}_{T},n)||^2\propto||\mathbf{h}^{0}_{T}-f_\theta(\mathbf{h}^{n}_{T},n)||^2
    \end{aligned}
\end{equation}
Therefore, we can train a neural network $f_\theta$ that directly predicts $\mathbf{h}^{0}_{T}$ based on $\mathbf{h}^{n}_{T}$, and use the posterior distribution $q(\mathbf{h}^{n-1}_{T}|\mathbf{h}^{n}_{T},\mathbf{h}^{0}_{T})$ to compute $\mu_{n}$ and $\mu_{\theta}$. Combining Eq. \ref{simple1} and Eq. \ref{simple2}, we can simplify Eq. \ref{VLB2} as:
\begin{equation}
\begin{aligned}
    \mathcal{L}^{''}_{\rm{VLB}} &=\mathbb{E}_{q(\mathbf{h}^{0:N}_{T} | v_T)}\Bigg[||{\mu}_{N}(\mathbf{h}^{N}_{T}, \mathbf{h}^{0}_{T}) ||^2+\sum_{n=2}^N||\mathbf{h}^{0}_{T}-f_\theta(\mathbf{h}^{n}_{T},n) ||^2\\&+||\mathbf{E}(v_T)-f_\theta(\mathbf{h}^{1}_{T},n)||^2-\log{p_{\hat{\phi}}(v_T|\mathbf{h}^{0}_{T})}\Bigg]
\end{aligned}
\end{equation}
Moreover, we notice that ${\mu}_{N}||(\mathbf{h}^{N}_{T}, \mathbf{h}^{0}_{T}) ||^2$ can be discarded because calculating the square of the fully noised representation $\mathbf{h}^{N}_{T}$ is unnecessary. Finally, we derive the simplified objective function for our framework:
\begin{equation}\label{L}
\begin{aligned}
\mathcal{L}&=\underbrace{\mathbb{E}_{q(\mathbf{h}^{1:N}_{T} | v_T)}\left[\left(\sum_{n=2}^N||\mathbf{h}^{0}_{T}-f_\theta(\mathbf{h}^{n}_{T},n)||^2\right)+||\mathbf{E}(v_T)-f_\theta(\mathbf{h}^{1}_{T},1)||^2\right]}_{\text{MSE Loss}}\\&+\underbrace{\mathbb{E}_{q(\mathbf{h}^{0}_{T} | v_T)}-\log{p_{\hat{\phi}}(v_T|\mathbf{h}^{0}_{T})}}_{\text{Rec Loss}}
\end{aligned}
\end{equation}
where the first and the second term calculate the MSE between the predicted hidden representation and the original hidden representation, the third term is the standard cross-entropy recommendation loss widely adopted in previous works such as BERT4Rec \cite{Sun2019bert} and DuoRec \cite{DuoRec}.
\subsection{Denoise Sequential Recommender}\label{DSR}
As we have mentioned above, we need to train a neural network $f_{\theta}$ to predict $\mathbf{h}^{0}_{T}$. In this section, we introduce the architecture of $f_{\theta}$, which we name it as the Denoise Sequential Recommender (DSR). DSR outputs the denoised hidden representation $f_\theta(\mathbf{h}^{n}_{T},n)$ based on the hidden representation of the whole sequence $\mathbf{H}^{n}$. To make DSR aware of the sequence order, we construct a learnable position embedding matrix $\mathbf{Y}\in\mathbb{R}^{T{\times}d}$ \cite{Devlin2019BERT} to represent all the $T$ timestamps. As DSR needs to handle the hidden representations at different noise levels, we also construct a diffusion-step embedding matrix to represent all the $N$ diffusion steps. We then sum up the sequence representation $\mathbf{H}^{n}$, the position embedding, and the diffusion-step embedding, to obtain $\widehat{\mathbf{H}}^{n}$:
\begin{equation}
    \widehat{\mathbf{H}}^{n}=[\mathbf{h}^{n}_{1}+\mathbf{y}_{1}+\mathbf{z}_{n},\mathbf{h}^{n}_{2}+\mathbf{y}_{2}+\mathbf{z}_{n},\cdots,\mathbf{h}^{n}_{t}+\mathbf{y}_{t}+\mathbf{z}_{n},\cdots,\mathbf{h}^{n}_{T}+\mathbf{y}_{T}+\mathbf{z}_{n}]
\end{equation}
where $\mathbf{y}_{t}\in\mathbb{R}^{d}(1{\leq}t{\leq}T)$ is the learnable position embedding vector for timestamp $t$, the diffusion-step embedding $\mathbf{z}_{n}\in\mathbb{R}^{d}(1{\leq}n{\leq}N)$ is obtained by the sinusoidal function \cite{vaswani2017transformer}:
\begin{equation}
\begin{aligned}
    &\mathbf{z}_{n}(2j)=\sin(n/10000^{2j/d}),\\ &\mathbf{z}_{n}(2j+1)=\cos(n/10000^{2j/d}),\thinspace\thinspace\thinspace0{\leq}{j}{\textless}d/2
\end{aligned}
\end{equation}
where $2j$ and $2j+1$ represent the dimension.

We then adopt the Transformer Encoder \cite{vaswani2017transformer} to process the sequence representation $\widehat{\mathbf{H}}^{n}$ and fetch the hidden representation at the last timestamp\footnote{Here $[-1]$ is mimicking the Python style of fetching the last element.}:
\begin{equation}\label{fthetahTn}
    f_\theta(\mathbf{h}^{n}_{T},n) = \text{TransformerEncoder}(\widehat{\mathbf{H}}^{n})[-1]
\end{equation}
\subsection{Efficient Training and Inference}\label{trainingandinference}
\subsubsection{Training}
A major challenge for training the diffusion models is the lengthened diffusion steps. If we decompose Eq. \ref{L} is a similar way as Eq. \ref{VLB}, we can see that Eq. \ref{L} is a combination of the MSE loss $\mathcal{L}_n$ plus the recommendation loss $\mathcal{L}_{\rm{Rec}}$:
\begin{equation}
\begin{aligned}
     &\mathcal{L} = \mathcal{L}_{N}  + \cdots + \mathcal{L}_{n} +\cdots + \mathcal{L}_{1}+\mathcal{L}_{\rm{Rec}}\\
     &\mathcal{L}_{n}=\mathbb{E}_{q(\mathbf{h}^{1:N}_{T} | v_T)}||\mathbf{h}^{0}_{T}-f_\theta(\mathbf{h}^{n}_{T},n)||^2,{\quad}2{\leq}n{\leq}N\\
     &\mathcal{L}_{1}=\mathbb{E}_{q(\mathbf{h}^{1:N}_{T} | v_T)}||\mathbf{E}(v_T)-f_\theta(\mathbf{h}^{1}_{T},1)||^2\\
    &\mathcal{L}_{\rm{Rec}}=\mathbb{E}_{q(\mathbf{h}^{0}_{T} | v_T)}-\log{p_{\hat{\phi}}(v_T|\mathbf{h}^{0}_{T})}\\
     \end{aligned}
\end{equation}
Previous works for sequential recommendation are usually optimized via the recommendation loss $\mathcal{L}_{\rm{Rec}}$, while our diffusion model requires additional calculation of the MSE loss $\mathcal{L}_n$ at different diffusion steps. However, if we calculate all the $\mathcal{L}_{1}, \mathcal{L}_{2}, \cdots, \mathcal{L}_{N}$ for each sequence $s_u$, then we have to compute all the model output $f_\theta(\mathbf{h}^{1}_{T},1),f_\theta(\mathbf{h}^{2}_{T},2),\cdots,f_\theta(\mathbf{h}^{N}_{T},N)$ and increase the time complexity by $N$ times. Given that the total diffusion steps required to achieve satisfactory performance is about 1000 \cite{DiffusionBeatGAN,DenoiseDiffusion}, increasing the time complexity by 1000 times can be unbearable for recommender systems with large-scale datasets. To address this problem and perform efficient training, we decide to only sample one diffusion step $n$ for each training sequence $s_u$ at each iteration\footnote{Under a typical setting with 20000 training sequences and 100 training epochs and 1000 diffusion steps, if we sample the diffusion steps uniformly, we will produce about $20000{\times}100/1000=2000$ samples for each diffusion step $n$. We believe that such amount of data will be sufficient for training DSR to handle the hidden representations at different noise levels.}. This modification allows us to calculate only one $\mathcal{L}_{n}$ plus $\mathcal{L}_{\rm{Rec}}$ for each training sequence $s_u$, thus achieving a similar time complexity with most of the deep-learning based sequential recommenders.

Moreover, in our experiments, we find out that the value of $\mathcal{L}_n$ at different diffusion steps vary significantly. When $n$ is relatively small, not much noise is added to $\mathbf{h}^{n}_{T}$, so DSR can easily denoise the hidden representation and hence yield a smaller $\mathcal{L}_n$; when $n$ approaches the total diffusion step $N$, denoising the hidden representation is challenging as $\mathbf{h}^{n}_{T}$ becomes approximately Gaussian noise, so $\mathcal{L}_n$ maintains a relatively large value. Intuitively, we should spend more diffusion steps on larger $\mathcal{L}_n$ and fewer diffusion steps on smaller $\mathcal{L}_n$, because training with harder samples will strengthen the denoising ability of DSR. Therefore, we adopt importance sampling \cite{ImprovedDiffusion} to encourage harder samples:
\begin{equation}\label{L_su}
    \mathcal{L}(s_u)=\mathbb{E}_{n\sim p_n}\left[\frac{\mathcal{L}_n}{p_n}\right]+\mathcal{L}_{\rm{Rec}}, \quad p_n \propto \sqrt{\mathbb{E}[\mathcal{L}_n^2]},\quad \sum_{n=1}^{N} p_n=1
\end{equation}
where $p_n$ is the probability of sampling the diffusion step $n$, $\mathcal{L}(s_u)$ is the weighted loss function. As $\mathbb{E}[\mathcal{L}_n^2]$ may change during training, we record the value of each $\mathcal{L}_n$ and update the record history dynamically. Based on the designs mentioned above, we summarize our training strategy in Algorithm \ref{alg:training}.
\begin{algorithm}
  \caption{Efficient Training} \label{alg:training}
  \begin{algorithmic}[1]
  \Require training dataset $\{s_u\}_{u=1}^{\mathcal{U}}$, batch size $S$, sequence length $T$, total diffusion steps $N$, noise schedules $\{\beta_{n}\}^{N}_{n=1}$
  \Ensure $\Theta=\{\theta,\phi,\hat{\phi}\}$
    \For{$\rm{Epoch} = 1,2,\cdots,\rm{MaxEpochs}$}
    \For{a minibatch $\{s_u\}_{u=1}^{S}$}
      \State Sample a diffusion step $n\sim p_n$
      \State Sample random Gaussian noise 
      $\epsilon \sim \mathcal{N}(0, \mathbf{I})$
      \State \textcolor{gray}{// \emph{Forward Process}}
      \State Calculate $\mathbf{h}^{0}_{t}(1{\leq}t{\leq}T)$ via ${q_{\phi}}(\mathbf{h}^{0}_{t}|v_t)$ (Eq. \ref{qphiht0vt})
      \State Calculate $\mathbf{h}^{n}_{t}(1{\leq}t{\leq}T)$ via $q(\mathbf{h}^{n}_{t}|\mathbf{h}^{0}_{t})$ (Eq. \ref{qhtnh0})
    \State \textcolor{gray}{// \emph{Reverse Process}}
      \State Predict $\mathbf{h}^{0}_{T}$ via $f_\theta(\mathbf{h}^{n}_{T},n)$ (Eq. \ref{fthetahTn})
      \State Predict $v_T$ via ${p_{\hat{\phi}}}(v_T|\mathbf{h}^{0}_{T})$ (Eq. \ref{pphihatvtht0})
    \State \textcolor{gray}{// \emph{Optimization}}
      \State Calculate $\mathcal{L}(s_u)$ via Eq. \ref{L_su}
      \State Take gradient descent step on $\nabla_\Theta \mathcal{L}(s_u)$
    \State \textcolor{gray}{// \emph{Dynamic Update}}
      \State Update the history value for $\mathcal{L}_n$ 
    \EndFor
    \EndFor
  \end{algorithmic}
\end{algorithm}
\subsubsection{Inference}
As we have mentioned in Section \ref{statement}, the sequential recommender predicts the probability of recommending the next item $v_{T+1}$ based on the whole sequence $[v_1,v_2,\cdots,v_{T}]$ during inference. As we do not know $v_{T+1}$, we have to start from diffusion step $N$ where the hidden representation of $v_{T+1}$ becomes full Gaussian noise.  We then perform the reverse process and denoise the hidden representation of $v_{T+1}$ until we get the prediction result.

In our implementation, we create an extra token $[unk]$ as a placehholder and append $[unk]$ to the end of the sequence $s_u$. Due to the limit of the maximum sequence length, the first item $v_1$ has to be dropped. Therefore, the inference user sequence $s_u$ becomes $[v_2,v_3,\cdots,v_T,[unk]]$, with the hidden representation $\mathbf{h}^{n}_{T}$ at the last position $T$ corresponding to the token $[unk]$. As we start the reverse process from diffusion step $N$, and according to Eq. \ref{gethn}, $\mathbf{h}^{N}_{T}=\epsilon$ becomes full Gaussian noise, so the extra token $[unk]$ will not interfere with the inference result. We can then perform the reverse process and get a series of denoised hidden representation $\mathbf{h}^{N-1}_{T},\mathbf{h}^{N-2}_{T},\cdots,\mathbf{h}^{0}_{T}$, and based on $\mathbf{h}^{0}_{T}$ we will know probability of each item being recommended by calculating ${p_{\hat{\phi}}}(v_T|\mathbf{h}^{0}_{T})$. We summarize the inference procedure in Algorithm \ref{alg:inference}.
\begin{algorithm}
  \caption{Inference of $v_{T+1}$ for user sequence $s_u$} \label{alg:inference}
  \begin{algorithmic}[1]
  \Require inference sequence $s_u=[v_2,v_3,\cdots,v_T,[unk]]$, sequence length $T$, total diffusion steps $N$
    \For{$n=N,N-1,\cdots,2$}
      \State Sample random Gaussian noise 
      $\epsilon \sim \mathcal{N}(0, \mathbf{I})$
       \State Calculate $\mathbf{h}^{0}_{t}(1{\leq}t{\leq}T)$ via ${q_{\phi}}(\mathbf{h}^{0}_{t}|v_t)$ (Eq. \ref{qphiht0vt})
      \State Calculate $\mathbf{h}^{N}_{t}(1{\leq}t{\leq}T)$ via $q(\mathbf{h}^{N}_{t}|\mathbf{h}^{0}_{t})$ (Eq. \ref{qhtnh0})
      \State $\mu_\theta(\mathbf{h}^{n}_{T})=\frac{\sqrt{\bar{\alpha}_{n-1}}\beta_n}{1-\bar{\alpha}_n}f_\theta(\mathbf{h}^{n}_{T},n) + \frac{\sqrt{\alpha_{n}}(1-\bar{\alpha}_{n-1})}{1-\bar{\alpha}_n} \mathbf{h}^{n}_{T}$
    \State Calculate $\mathbf{h}^{n-1}_{T}$ via $p_{\theta}(\mathbf{h}^{n-1}_{T}|\mathbf{h}^{n}_{T})$ (Eq. \ref{pthetahn1hn})
    \EndFor
    \State Predict $\mathbf{h}^{0}_{T}$ via $f_\theta(\mathbf{h}^{1}_{T},1)$ (Eq. \ref{fthetahTn})
    \State Obtain probability distribution $p\gets{p_{\hat{\phi}}}(v_T|\mathbf{h}^{0}_{T})$ (Eq. \ref{pphihatvtht0})
   \State\Return $p$ 
  \end{algorithmic}
\end{algorithm}

However, a major problem with Algorithm \ref{alg:inference} is the lengthened diffusion steps. As we have to start from $\mathbf{h}^{N}_{T}$ until we reach $\mathbf{h}^{0}_{T}$, the time complexity of the inference procedure is increased by $N$ times. To address this issue, we notice that $f_\theta$ is trained to directly predict $\mathbf{h}^{0}_{T}$ based on any $\mathbf{h}^{n}_{t}(1{\leq}n{\leq}N)$, so it is able to directly predict $\mathbf{h}^{0}_{T}$ from $\mathbf{h}^{N}_{T}$ without the need of the intermediate diffusion steps. Furthermore, we find out the recommendation performance can be improved by averaging the predictions yielded from different random seeds\footnote{During inference, deterministic approaches such as SASRec and BERT4Rec will generate the same recommendation result even if we use different random seeds. By comparison, a good property of diffusion models is that we can sample different Gaussian noise $\epsilon$ and generate different recommendation results.}. This is probably because different recommendation results can consider different situations of behavior uncertainty, and aggregating the results under different situations adds diversity to the overall recommendation results (Section~\ref{uncertaintydiversity}). Combining these two techniques, we design an efficient inference procedure in Algorithm \ref{alg:efficientinference}.
\begin{algorithm}[htbp]
  \caption{Efficient inference of $v_{T+1}$ for user sequence $s_u$} \label{alg:efficientinference}
  \begin{algorithmic}[1]
  \Require inference sequence $s_u=[v_2,v_3,\cdots,v_T,[unk]]$, sequence length $T$, total diffusion steps $N$, random seeds list $seeds$
    \For{$i=1,2\cdots,|seeds|$}
    \State Fix random seed as $seeds[i]$
      \State Sample random Gaussian noise 
      $\epsilon \sim \mathcal{N}(0, \mathbf{I})$
       \State Calculate $\mathbf{h}^{0}_{t}(1{\leq}t{\leq}T)$ via ${q_{\phi}}(\mathbf{h}^{0}_{t}|v_t)$ (Eq. \ref{qphiht0vt})
      \State Calculate $\mathbf{h}^{N}_{t}(1{\leq}t{\leq}T)$ via $q(\mathbf{h}^{N}_{t}|\mathbf{h}^{0}_{t})$ (Eq. \ref{qhtnh0})
      \State Predict $\mathbf{h}^{0}_{T}$ via $f_\theta(\mathbf{h}^{N}_{T},N)$ (Eq. \ref{fthetahTn})
      \State Obtain probability distribution $p_i\gets{p_{\hat{\phi}}}(v_T|\mathbf{h}^{0}_{T})$ (Eq. \ref{pphihatvtht0})
    \EndFor
   \State\Return $\frac{1}{|seeds|}\sum_{i=1}^{|seeds|}p_i$ 
  \end{algorithmic}
\end{algorithm}

Algorithm \ref{alg:efficientinference} greatly saves computational cost by directly predicting $\mathbf{h}^{0}_{T}$ from $\mathbf{h}^{N}_{T}$. As we normally set the size of random seeds list as a small number (i.e., 10 by default), the additional computational cost incurred by using different random seeds can be minimal.
\section{EXPERIMENT}\label{experiment}

\subsection{Settings}
\subsubsection{Dataset.}
We conduct experiments on three public benchmark datasets collected from two online platforms. The Amazon dataset comprises users' reviews on products from different categories, and we select two subcategories, \textbf{Beauty} and \textbf{Toys}, for our experiments. The MovieLens dataset \cite{movielens} is also a popular benchmark dataset containing users' ratings on movies, and we use the version (\textbf{ML-1M}) with about 1 million ratings.

We follow previous works \cite{kang18attentive,Xu2020Contrastive,Sun2019bert} to preprocess the datasets. We treat all interactions as implicit feedbacks, discard users or items related with less than 5 interactions, and construct user sequences by sorting the users' interactions in the chronological order. The preprocessed dataset statistics are presented in Table \ref{dataset}.
\begin{table}
    \centering
    \caption{\textbf{Dataset statistics after preprocessing}}
    \label{dataset}
  \setlength{\tabcolsep}{5pt}{
    \begin{tabular}{cccccc}
    \toprule
         Datasets&\#users&\#items&\#actions&avg.length&sparsity\\\midrule
         Beauty&22,363&12,101&198,502&8.9&99.93\%\\
         Toys&19,412&11,924&167,597&8.6&99.93\%\\
         ML-1M&6,040&3,953&1,000,209&163.5&95.81\%\\\bottomrule
    \end{tabular}}
\end{table}
\subsubsection{Evaluation Metrics} To evaluate the performance of sequential recommenders, we choose the score of top-$K$ Hit Ratio (HR@$K$) and Normalized Discounted Cumulative Gain (NDCG@$K$) as the evaluation metrics, where $K\in\{5,10,20\}$. The \textit{leave-one-out} evaluation strategy is adopted, leaving out the last item for test, the second-to-last item for validation, and the rest for training. As sampled metrics might lead to unfair comparisons \cite{krichene2020sampled}, we rank the prediction results on the whole dataset without negative sampling.
\subsubsection{Baselines} To provide a comprehensive evaluation, we compare the performance of our approach with simple recommendation methods, standard sequential methods, 
 and generative sequential recommendation methods. As contrastive learning is becoming popular in sequential recommendation, we also compare our approach with two representative contrastive sequential methods. A brief introduction on the baseline methods is presented as follows:
\begin{itemize}
[leftmargin =  8pt,topsep=1pt]
\item\textbf{Simple recommendation methods} include PopRec and BPR-MF. PopRec is a simple recommendation strategy that always recommends the most popular items to all users. BPR-MF \cite{BPR} is a matrix factorization \cite{factorizationmachines} model with Bayesian Personalized Ranking (BPR) optimization.
\item\textbf{Standard sequential methods} adopt a neural network as sequence encoder to capture the correlations between the items. For example, GRU4Rec \cite{srnn2016} utilizes RNN as sequence encoder, SASRec \cite{kang18attentive} and BERT4Rec \cite{Sun2019bert} adopt Transformers as sequence encoder, STOSA \cite{STOSA} adopts a stochastic Transformer with Wasserstein self-attention as sequence encoder.
\item\textbf{Generative sequential methods} adopt generative models such as VAE and GAN to capture the uncertainty of user behaviors and alleviate exposure bias. For example, ACVAE \cite{ACVAE} adopts an adversarial and contrastive variational autoencoder to learn more personalized and salient characteristics, VSAN \cite{VSAN} utilizes a variational self-attention network to characterize the uncertainty of user preferences, MFGAN \cite{MFGAN} employs a multi-factor generative adversarial network to consider information of various factors and alleviate exposure bias.
\item\textbf{Contrastive sequential methods} enhance sequential recommendation models with contrastive learning to mitigate the data sparsity issue. For example, CL4SRec \cite{Xu2020Contrastive} proposes data augmentation strategies with contrastive learning objectives for sequential recommendation, DuoRec \cite{DuoRec} proposes both supervised and unsupervised contrastive learning strategies to alleviate the representation degeration problem in sequential recommendation.
\end{itemize}
\subsubsection{Implementation} PopRec and BPR-MF are implemented on public resources. For GRU4Rec, SASRec, BERT4Rec, STOSA, ACVAE, MFGAN and DuoRec, we use the codes provided by their authors. We implement VSAN and CL4SRec in PyTorch \cite{pytorch}. We follow the instructions from the original papers to set and tune the hyper-parameters.

We also implement our framework in PyTorch \cite{pytorch}. We adopt a standard Transformer encoder with 2 layers and 2 attention heads as the backbone of DSR. We set the batchsize as 256, hidden dimensionality as 256, sequence length as 50 for all the datasets. We tune the dropout ratio on each dataset and the optimal dropout ratio is set as 0.3 for Beauty and Toys, 0.2 for ML-1M. We use the Adam \cite{Adam} optimizer with a learning rate of $0.001$. We use the default setting with the sqrt noise schedule \cite{li2022diffusion} and 1000 diffusion steps \cite{DenoiseDiffusion}, and other variants are investigated in Section \ref{FurtherAnalysis}. We train our model for 50 epochs and select the checkpoint with the best NDCG@10 score on the validation set for test. 
\begin{table*}
  \centering
     \caption{Overall performance of different methods for sequential recommendation. The best score and the second-best score in each row are bolded and underlined, respectively. The last column indicates improvements over the best baseline method.}

  \setlength{\tabcolsep}{1.8pt}{

    \begin{tabular}{c|l|cc|cccc|ccc|cc|c|c}
\toprule
Dataset&Metric&PopRec&BPR-MF&GRU4Rec&SASRec &BERT4Rec&STOSA&ACVAE&VSAN&MFGAN&CL4SRec&DuoRec&DiffRec&Improv. \\
\midrule   \multirow{6}[0]{*}{Beauty} 
&HR@5&0.0072& 0.0120&0.0206&0.0371&0.0370&0.0460&0.0438&0.0475&0.0382&0.0396&\underline{0.0541}&\textbf{0.0629}& 16.27\%\\
&HR@10  &0.0114& 0.0299&0.0332&0.0592&0.0598&0.0659&0.0690&0.0759 &0.0605&0.0630&\underline{0.0825}&\textbf{0.0847}& 2.67\%\\
&HR@20  &0.0195& 0.0524&0.0526&0.0893&0.0935&0.0932&0.1059&0.1086&0.0916&0.0965&\underline{0.1102}&\textbf{0.1125}& 2.09\%\\
&NDCG@5 & 0.0040&0.0065&0.0139&0.0233&0.0233&0.0318&0.0272&0.0298&0.0254&0.0232&\underline{0.0362}&\textbf{0.0447}& 23.48\%\\
&NDCG@10&0.0053& 0.0122&0.0175&0.0284&0.0306&0.0382&0.0354&0.0389&0.0310&0.0307&\underline{0.0447}&\textbf{0.0518}& 15.59\%\\
&NDCG@20&0.0073&  0.0179&0.0221&0.0361&0.0391&0.0451&0.0453&0.0471&0.0405&0.0392&\underline{0.0531}&\textbf{0.0588}& 10.73\%\\
\midrule  \multirow{6}[0]{*}{Toys} 
&HR@5 &0.0065&0.0122&0.0121& 0.0429  & 0.0371 & \underline{0.0563} &0.0457&0.0481&0.0395& 0.0503 &0.0539& \textbf{0.0647}& 14.92\%\\
&HR@10&0.0090&0.0197&0.0184&0.0652&0.0524&\underline{0.0769}&0.0663&0.0719&0.0641&0.0736&0.0744& \textbf{0.0834}  & 8.45\%\\
&HR@20&0.0143& 0.0327 &0.0290&0.0957&0.0760&0.1006&0.0984&0.1029&0.0892&0.0990&\underline{0.1056}&\textbf{0.1088}  & 3.03\%\\
&NDCG@5&0.0044& 0.0076&0.0077&0.0248&0.0259&\underline{0.0393}&0.0291&0.0286&0.0257&0.0264&0.0340&\textbf{0.0474} & 20.61\%\\
&NDCG@10&0.0052&0.0100&0.0097&0.0320&0.0309&\underline{0.0460}&0.0364&0.0363&0.0328&0.0339&0.0406&\textbf{0.0534} & 16.09\%\\
&NDCG@20&0.0065& 0.0132&0.0123&0.0397&0.0368&\underline{0.0519}&0.0432&0.0441&0.0381&0.0404&0.0472&\textbf{0.0598} & 15.22\%\\
\midrule    
  \multirow{6}[0]{*}{ML-1M} 
&HR@5&0.0078&0.0164&0.0806&0.1078&0.1308&0.1230&0.1356&0.1220&0.1275&0.1142&\underline{0.1679}&\textbf{0.2022}&20.43\%\\
&HR@10&0.0162& 0.0354&0.1344& 0.1810 & 0.2219 &0.1889& 0.2033 & 0.2015&0.2086&0.1815&\underline{0.2540}& \textbf{0.2900}&14.17\%\\
&HR@20& 0.0402& 0.0712&0.2081&0.2745&0.3354&0.2724&0.3085&0.3003&0.3166&0.2818&\underline{0.3478}&\textbf{0.3923}&12.79\%\\
&NDCG@5&0.0052& 0.0097&0.0475&0.0681&0.0804&0.0810&0.0837&0.0751&0.0778&0.0705&\underline{0.1091}&\textbf{0.1374}&25.94\%\\
&NDCG@10&0.0079& 0.0158&0.0649&0.0918&0.1097&0.1021&0.1145&0.1007&0.1040&0.0920&\underline{0.1370}&\textbf{0.1657}&20.95\%\\
&NDCG@20& 0.0139&0.0248&0.0834&0.1156&0.1384&0.1231&0.1392&0.1257&0.1309&0.1170&\underline{0.1607}&\textbf{0.1916}&19.23\%\\
\bottomrule\end{tabular}}
  \label{performancecomparison}
\end{table*}
\subsection{Overall Performance Comparisons}
Table \ref{performancecomparison} compares the overall performances of our model and other baseline methods on three public benchmark datasets. Based on the results, we can see that:
\begin{itemize}
[leftmargin =  8pt,topsep=1pt]
\item DiffRec consistently outperforms all the baseline methods on both sparse dataset (e.g., Beauty, Toys) and dense dataset (e.g., ML-1M). We attribute the performance improvement to these factors: (1) DiffRec treats item representations as probability distributions and requires sampling from random Gaussian noises, which can mimic the uncertainty of user behaviors in real-world situations; (2) Instead of using a neural network to estimate the mean and variance, DiffRec adopts a predefined Markov chain to control the forward process, which can generate high-quality samples with less information loss. Noising the target item instead of the whole sequence also helps to retain essential information about user preferences; (3) The inference strategy adds diversity to the recommendation results and further improves performance. 
\item Among simple recommendation methods and standard sequential methods, STOSA achieves the best performance. This is probably because its stochastic Transformer is not a deterministic architecture and can capture the uncertainty of user behaviors. Moreover, generative sequential recommendation methods achieve similar performance with STOSA and outperform other standard sequential methods. This is probably because generative models introduce random sampling processes to estimate the mean and variance of item representations, which can also capture the uncertainty of user behaviors.
\item Among all the baseline methods, contrastive sequential methods generally perform better than other methods. This is probably because contrastive learning enriches the training data by constructing more augmented sequences, which are helpful for alleviating the data sparsity issue. It is also worth noting that DiffRec can outperform contrastive sequential methods, although it is not yet equipped with any contrastive learning module. This further verifies the effectiveness of our approach.
\end{itemize}
\subsection{Further Analysis}\label{FurtherAnalysis}
\subsubsection{Influences of Noise Schedules}\label{influencesofnoiseschedule}
In this section, we investigate the influences of different noise schedules. We consider three kinds of noise schedules: 
\begin{itemize}
[leftmargin =  8pt,topsep=1pt]
\item \textbf{The sqrt noise schedule} \cite{li2022diffusion} is our default setting. It quickly rises the noise levels at the first few diffusion steps and gradually slows down the injection of diffusion noises at the latter diffusion steps. It is defined as: $\bar{\alpha}_n=1-\sqrt{n/N+0.0001}$.
\item \textbf{The cosine noise schedule} \cite{ImprovedDiffusion} smoothly increases diffusion noises using a cosine function to prevent sudden changes in the noise level. It is defined as: $\bar{\alpha}_n=\frac{g(n)}{g(0)},g(n)=\cos({\frac{n/N+0.008}{1+0.008}\cdot\frac{\pi}{2}})^{2}$.
\item \textbf{The linear noise schedule} \cite{DenoiseDiffusion} is originally proposed for image generation. It increases $\beta_n$ linearly from $\beta_1=10^{-4}$ to $\beta_N=0.02$.
\end{itemize}

\begin{figure}
\centering
{\includegraphics[ width=\linewidth]{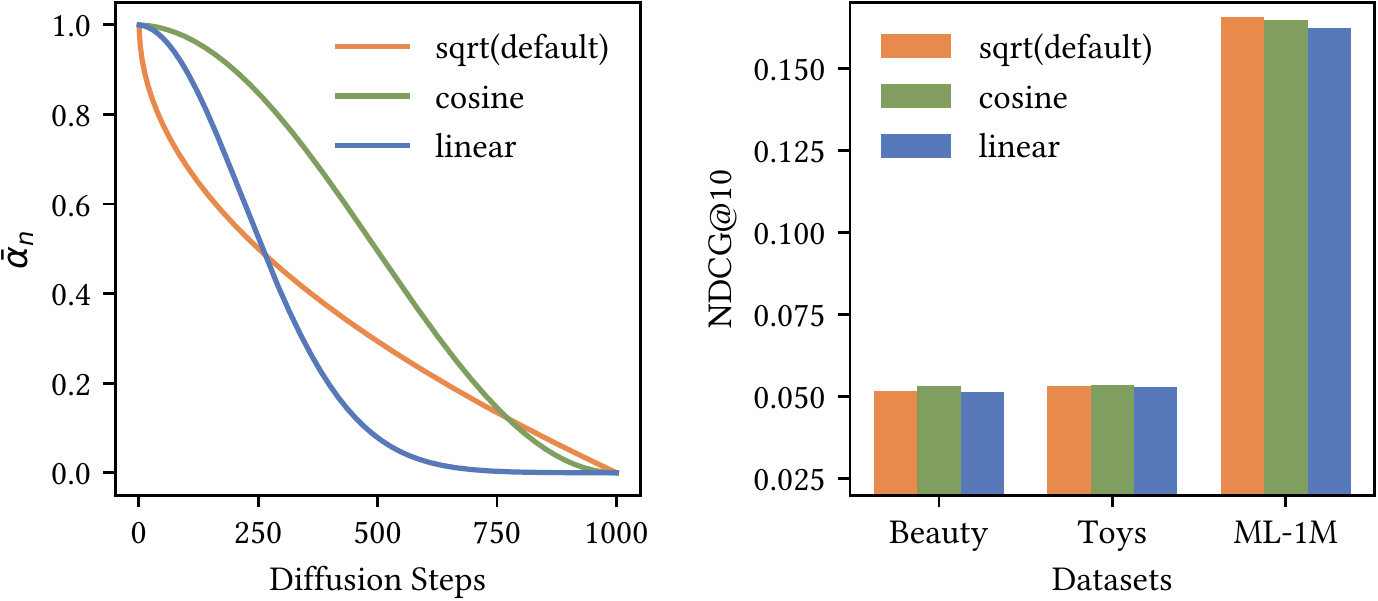}}
\caption{Influences of different noise schedules. The left figure is a visualization of $\bar{\alpha}_{n}$ under different noise schedules, and the right figure shows the performances (NDCG@10) on three datasets using different noise schedules.}\label{noiseschedulecompare}
\end{figure}

To compare their differences, we provide a visualization of different noise schedules and compare the performances on three dataset using different noise schedules. Since $\bar{\alpha}_n$ and $\beta_n$ are mutually convertible\footnote{$\bar{\alpha}_n$ is defined as $\bar{\alpha}_n = \prod_{i=1}^n 1-\beta_i$ and we note that $\beta_n=1-\frac{\bar{\alpha}_n}{\bar{\alpha}_{n-1}}$}, we plot the value of $\bar{\alpha}_n$ with 1000 diffusion steps (default setting), and lower value of $\bar{\alpha}_n$ means higher noise level. All the results are presented in Figure \ref{noiseschedulecompare}.

We can see that the performances do not vary much under different noise schedules. The default noise schedule, sqrt noise schedule, generally performs well, achieving the best performance on Toys and ML-1M dataset and the second best performance on Beauty dataset. The cosine noise schedule shows similar performance with the sqrt schedule, while the linear noise schedule shows performance slightly lower than the sqrt or the cosine noise schedule.
\subsubsection{Influences of Diffusion Steps} 
\begin{figure}
\centering
{\includegraphics[width=.495\linewidth]{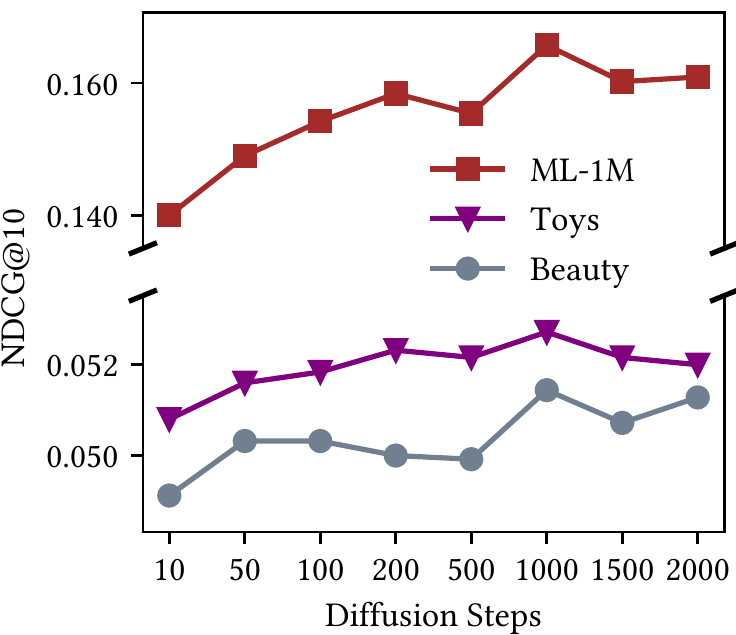}}
{\includegraphics[width=.495\linewidth]{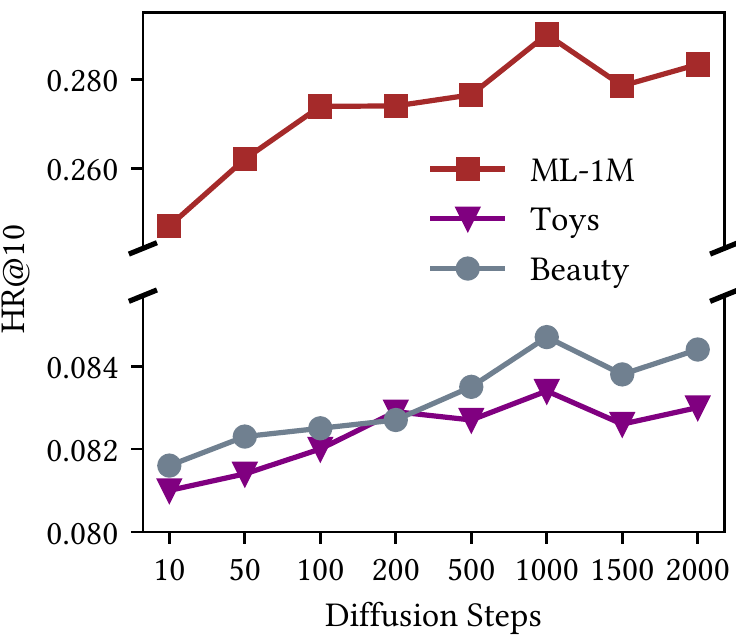}}
\caption{Performance comparison (NDCG@10, HR@10) w.r.t. different diffusion steps.}\label{influencesofdiffusionsteps}
\end{figure}
Generally speaking, diffusion models require a large number of diffusion steps $N$ to achieve satisfactory performance. This is probably because a large $N$ allows for better approximation of the Gaussian diffusion process \cite{NonequilibriumThermodynamics}. To investigate the influences of diffusion steps, we train DiffRec using different diffusion steps ranging from 10 to 2000 (the default number of diffusion steps is set as 1000 following \cite{DenoiseDiffusion,NonequilibriumThermodynamics}) and report the performances in Figure \ref{influencesofdiffusionsteps}. We can see that DiffRec generally performs well with large diffusion steps (e.g., 1000, 1500, 2000), and reaches the best performance with 1000 diffusion steps. However, we observe a performance decrease when we use fewer diffusion steps, which indicates the necessity of using larger diffusion steps.
\subsubsection{Ablation Study on Sequence Embeddings}
\begin{figure}
\centering
{\includegraphics[width=.495\linewidth]{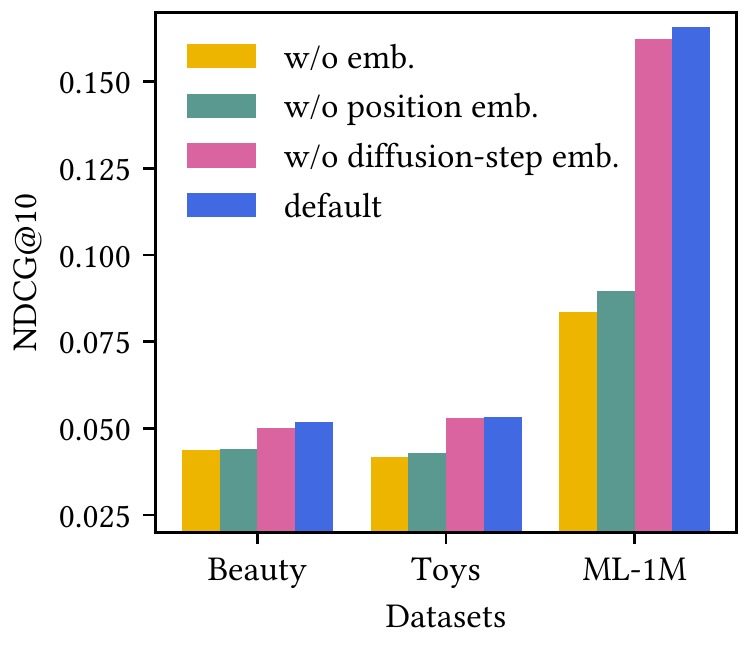}}
{\includegraphics[width=.495\linewidth]{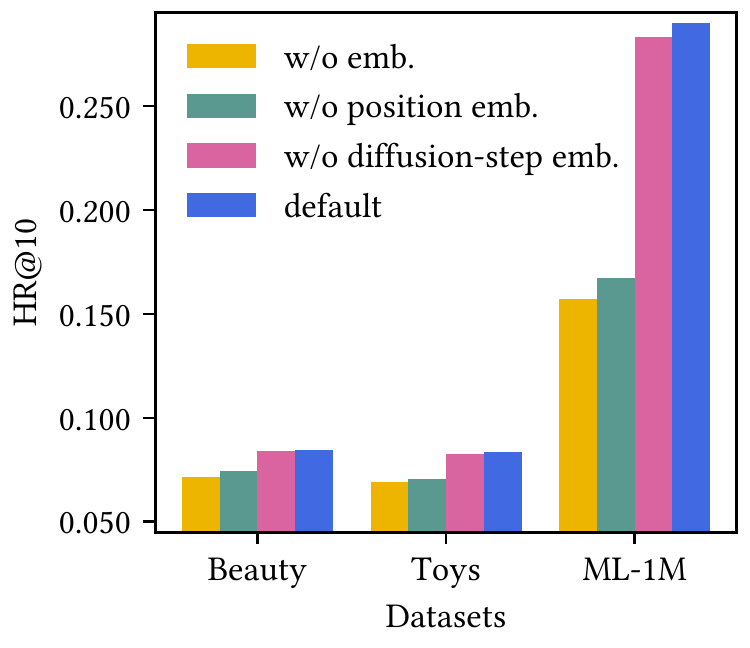}}
\caption{Ablation study (NDCG@10, HR@10) on sequence embeddings.}\label{ablationstudysequenceembeddings}
\end{figure}
Our framework requires a Denoise Sequential Recommender (DSR) to predict the original data, which is a Transformer encoder with position embedding and diffusion-step embedding. To verify the effectiveness of each embedding, we design three kinds of variants: (1) DSR without any embedding; (2) DSR without position embedding; (3) DSR without diffusion-step embedding. We compare the performances of these variants with the default setting, which is equipped with both the position embedding and the diffusion-step embedding. From the experiment results in Figure \ref{ablationstudysequenceembeddings}, we can see that all the variants exhibit worse performance than the default setting, and DSR without any embedding exhibits the worst performance. This phenomenon indicates that it is necessary to make DSR aware of both the sequence order and the diffusion step. While previous works \cite{kang18attentive,Sun2019bert} have verified that the learnable position embedding plays a crucial role in representing the sequence order, we also find out that the diffusion-step embedding is critical to representing the diffusion-step information. Without the diffusion-step information, DSR is unaware of the noise level of the hidden representations, thus achieving worse performance.
\subsubsection{Effectiveness of Noising Strategy}\label{EffectivenessofNoisingStrategies}
\begin{figure}
\centering
{\includegraphics[width=.495\linewidth]{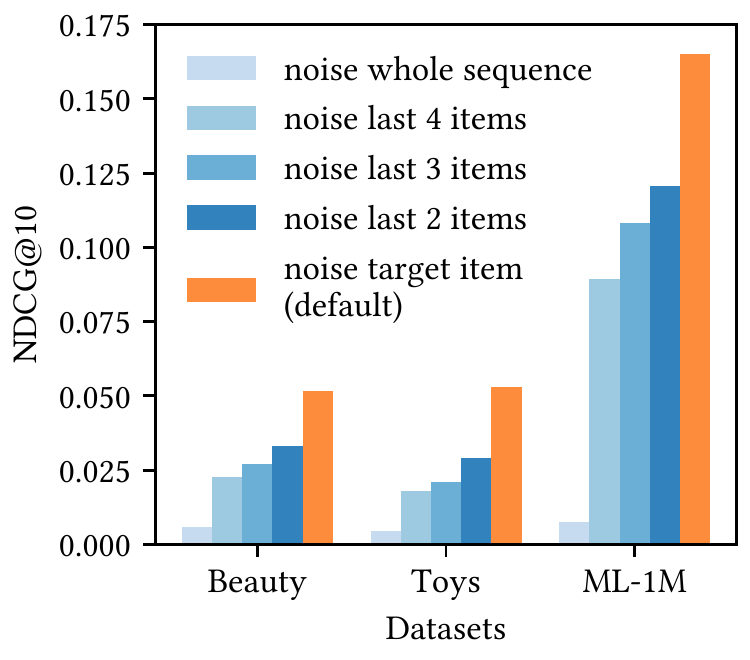}}
{\includegraphics[width=.495\linewidth]{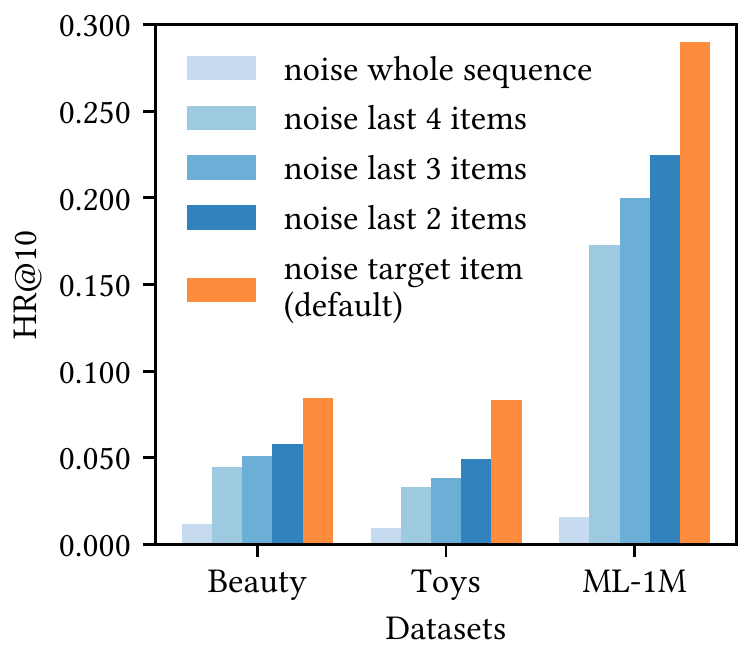}}
\caption{Performance comparison (NDCG@10, HR@10) w.r.t different noising strategies.}\label{noisingstrategy}
\end{figure}
Different from general diffusion models that noise the whole data, our framework adopts a different noising strategy that only noises the target item instead
of the whole sequence. To verify the effectiveness of such noising strategy, we design four kinds of variants: (1) noising the whole sequence; (2) noising the last 4 item; (3) noising the last 3 item; (4) noising the last 2 items. We compare the performances of these variants with the default setting, which only noises the target item (the last item). From the experiment results in Figure~\ref{noisingstrategy}, we can see that the default noising strategy achieves the best performance, while other variants achieve worse performance. Moreover, the performance steadily decreases as more items are noised, and noising the whole sequence achieves the worst performance. This is probably because sequential recommendation is a \emph{next item prediction task}, so applying noises to items other than the target item will corrupt essential information about user's preference, thus hurting performance. However, noising the last $K$ items might be useful when the task becomes predicting the next $K$ items (e.g., next-basket recommendation), which is an interesting direction for future research.
\subsubsection{Uncertainty \& Diversity}
\begin{figure*}
    \centering
\includegraphics[ width=\linewidth, trim=0 258.5 83 0,clip ]{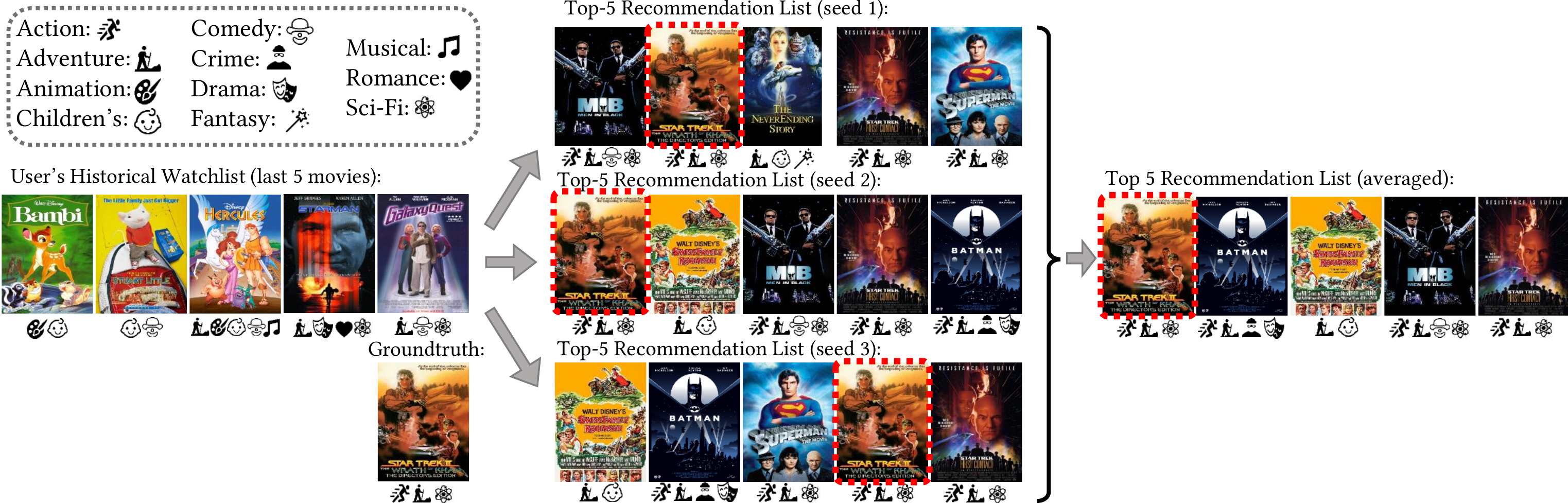}
    \caption{An example of user's watchlist and recommended movies. The groundtruth movie is marked in a red box. For brevity we only plot the top-5 recommendation list of 3 different random seeds.}
    \label{casestudy}
\end{figure*}
\begin{figure}
    \centering
{\includegraphics[width=.49\linewidth]{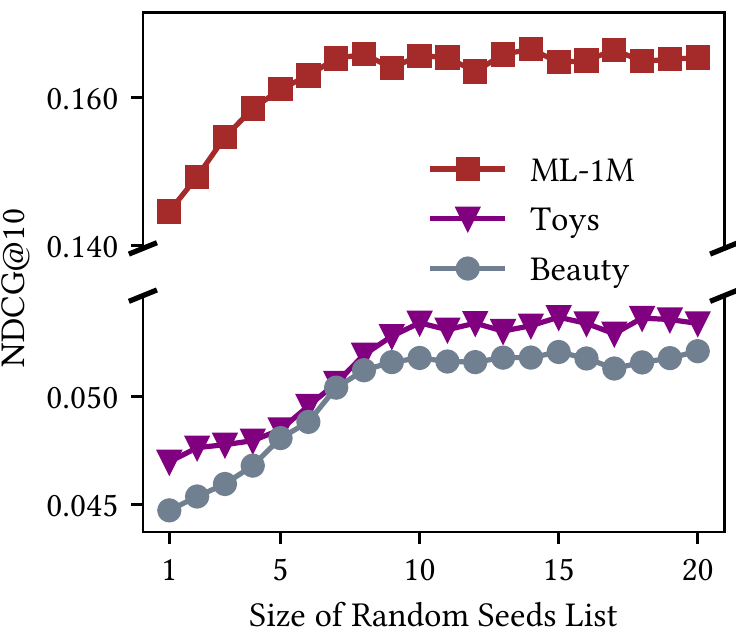}}
{\includegraphics[width=.49\linewidth]{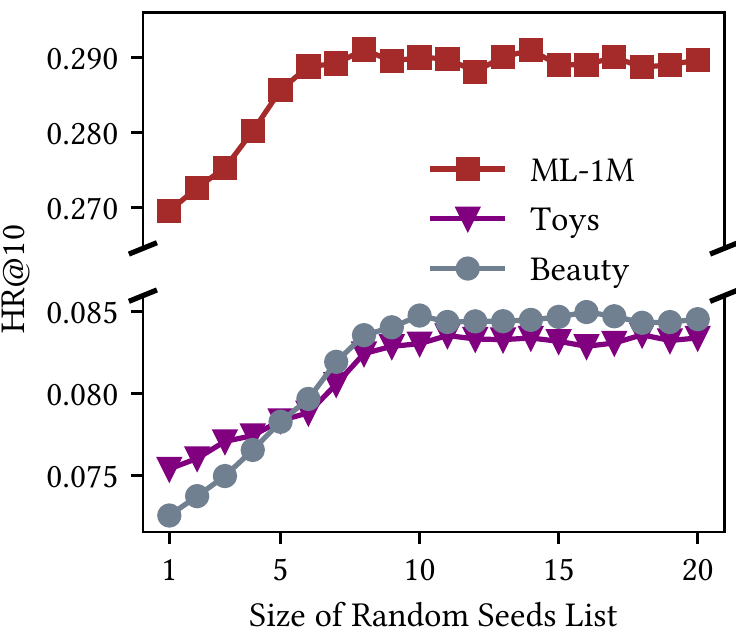}}
    \caption{Performance comparison (NDCG@10, HR@10) w.r.t different numbers of random seeds.}
    \label{inferenceseed}
\end{figure}
\label{uncertaintydiversity}
In this section, we aim to illustrate how DiffRec can estimate the uncertainty of user behaviors and add diversity to recommendation results. First, we provide a case study from the ML-1M dataset. The ML-1M dataset contains rich side information (i.e., movies' genres) that might be helpful for understanding user behaviors. We randomly select a user \#5200 from the ML-1M dataset and plot this user's watchlist as well as the recommendation lists in Figure~\ref{casestudy}. From the user's historical watchlist we can infer that this user favours movies with Action, Adventure and Sci-fi genres, so the recommender should recommend relevant movies with these genres. However, there are several movies (e.g., Men in Black and Star Trek) that satisfy the requirements, and the users may choose different movies under different situations due to behavior uncertainty. In such cases, using different random seeds to generate different recommendation results (Algorithm~\ref{alg:efficientinference}) can consider different situations of behavior uncertainty and improve the diversity of recommendation results. In Figure~\ref{casestudy}, we can see that the top-5 recommendation list of different random seeds are all relevant (with Action, Adventure and Sci-fi genres) but different movies, and averaging all the recommendation lists will generate a more diverse and accurate recommendation result.

Next, we quantitatively analyze to what extent considering different situations of behavior uncertainty contributes to performance improvement. We set the size of random seeds list from 1 to 20 and report the performances in Figure~\ref{inferenceseed}. We can see that the performances steadily increase as we use more random seeds. This verifies that considering more recommendation results from different random seeds will be helpful for performance improvement. However, performances reach a plateau when there are sufficient numbers of random seeds, indicating that the effectiveness of considering different recommendation results also has an upper limit.

\section{CONCLUSION} 
In this paper, we present sequential recommendation with diffusion models (DiffRec), a diffusion-model based sequential recommender capable of making high-quality recommendations with efficient training and inference cost. To adapt diffusion models for sequential recommendation, we modify the forward process and the reverse process to enable the processing of the discrete recommendation data. Objective function is then derived based on the modified diffusion process and a denoise sequential recommender is designed to fulfill the objective function. We also design an efficient training strategy and an efficient inference strategy to save computational cost and improve recommendation diversity. Extensive experiment results on three benchmark datasets verify the effectiveness of our approach.



\bibliographystyle{ACM-Reference-Format}
\balance
\bibliography{sample-base}


\end{document}